\documentclass[prb,aps,twocolumn]{revtex4}

\def\pr{Phys. Rev.\ }
\def\jour#1#2#3#4{{#1}{\bf #2}, #3 (#4)}
\def\tit#1#2#3#4#5{{#1} {\bf #2}, #3 (#4)}

\def\prl{Phys.\ Rev.\ Lett.}
\def\pr{Phys.\ Rev.}
\def\prb{Phys.\ Rev.\ B}

\def\zpb{Z.\ Phys.\ B}

\def\ijmpb{Int.\ J.\ Mod.\ Phys.\ B}
\def\jmp{J. Math.\ Phys.}
\newcommand{\bfx}{{\mathbf{x}}}
\newcommand{\bfy}{{\mathbf{y}}}
\newcommand{\bfk}{{\mathbf{k}}}
\newcommand{\kx}{k_x}
\newcommand{\ky}{k_y}

\newcommand{\bea}{\begin{eqnarray}}
\newcommand{\eea}{\end{eqnarray}}
\newcommand{\wtpsi}{\widetilde{\psi}}
\newcommand{\wtchi}{\widetilde{\chi}}

\usepackage{amsmath}  
\usepackage{epsfig}
\usepackage{bm}

\begin{document}


\title{
Classical dimers on the triangular lattice
}

\author{P. Fendley,$^1$
R. Moessner,$^2$ S.\,L. Sondhi$^3$
}

\affiliation{\medskip
$^1$ Department of Physics, University of Virginia,\\
Charlottesville, VA 22904-4714, USA \medskip\\
$^2$ Laboratoire de Physique Th\'eorique de l'Ecole Normale
Sup\'erieure, CNRS-UMR8541, Paris, France\medskip\\
$^3$ Department of Physics, Princeton University,
Princeton, NJ 08544, USA \medskip
} 

\date{original date: June 2, 2002}

\begin{abstract}
We study the classical hard-core dimer model on the triangular lattice.
Following Kasteleyn's fundamental theorem on planar graphs, this problem
is soluble using Pfaffians. This model is particularly interesting 
for, unlike the dimer problems on the bipartite square and hexagonal 
lattices, its correlations are short ranged with a correlation length
of less than one lattice constant.
We compute the dimer-dimer and monomer-monomer correlators,
and find that the model is deconfining: the
monomer-monomer correlator falls off exponentially
to a constant value $\sin(\pi/12)/\sqrt{3}\approx .1494\dots$,
only slightly below the nearest-neighbor value of $1/6$.
We also consider the anisotropic triangular lattice
model in which the square lattice is perturbed by diagonal
bonds of one orientation and small fugacity. 
We show that the model becomes non-critical immediately
and that this perturbation is equivalent to adding a mass term to each of two
Majorana fermions that are present in the long wavelength limit of the square
lattice problem.
\end{abstract}


\maketitle

\section{Introduction}
The study of classical dimer models -- the statistical mechanics of
hardcore dimer coverings of graphs -- has a venerable 
history. These models have been of interest as direct 
representations of the physics, e.g.\ diatomic molecules on a lattice 
but even more because of their equivalence to various other
statistical mechanical problems; for example, the two-dimensional
Ising model can be reformulated as a dimer model on a special
lattice.\cite{kasteleyn,fishdimis}
The reduction to dimer form is advantageous
in that a wide class of dimer models, those on planar graphs with 
independent fugacities on bonds, are soluble by Pfaffians following
a theorem of Kasteleyn.\cite{kasteleyn,NYB} This technique has been
applied to compute a variety of correlation functions, in the dimer
model on the square lattice \cite{Fisher63}
and in the Ising model.\cite{MW}

The interest in dimer models has received new and very different impetus 
following the
discovery of high-temperature superconductivity. Following Anderson's
proposal\cite{Anderson87} that the superconducting state evolves out
of a liquid of singlet (valence) bonds, Rokhsar and Kivelson proposed
a {\it quantum} dimer model \cite{Rokhsar88} to describe this so-called
short-range resonating valence bond (RVB) physics. 
The Hilbert space of their model
consists of all pairings of the spins on the square lattice into
singlet bonds; these can simply be labeled by  hardcore dimer
coverings. The quantum dynamics provides off-diagonal matrix elements
between these coverings. This idea can be generalized to other
lattices.

The properties of the classical dimer model enter the solution of such
quantum dimer models in two ways. First, trivially but usefully, the
infinite temperature statics of the quantum problem are given
precisely by the classical problem with equal fugacities. Second,
non-trivially and even more usefully, quantum dimer models generically
exhibit an ``RK'' point -- which generalizes Rokhsar and Kivelson's
construction on the square lattice -- where the quantum wavefunction
is an equal amplitude superposition of dimer coverings which is the
paradigmatic RVB form. Static, dimer diagonal, ground state
correlations at the RK point are then again given by the correlations
of the classical dimer model.\cite{Rokhsar88,kenyon}
For the square lattice,
the known results on the classical problem \cite{Fisher63} showed that
the quantum dimer model was critical at the RK point, which thus
turned out to be an isolated critical point between two solid phases,
rather than a representative of an RVB phase. Similar behavior occurs
on the honeycomb lattice, where the classical model is equivalent to
the five-vertex model on the square lattice.\cite{Wu}

In recent work, two of the present authors have shown that the
triangular lattice leads to a a different outcome \cite{MStrirvb} with
an RVB {\it phase}, including the RK point, characterized by liquid
correlations. It was also noted that the model exhibits ``topological
order'' in the sense of Wen\cite{wen} (ground state degeneracies on
closed surfaces in the absence of symmetry breaking) characteristic of
an Ising gauge theory.\cite{MStrirvb,duality} As part of this work it
was necessary to solve the isotropic classical dimer model on the
triangular lattice and show that the dimer correlations were
short-ranged. As this does not appear to have been done previously,
with the exception of results for the thermodynamic limit
entropy,\cite{nagle,gaunt,samuel,mila,wolffzitt} we have carried out a
fuller analysis of the problem in this paper.

Two aspects of this expanded analysis are noteworthy. First, we show
that monomers are deconfined on the triangular lattice. A monomer is a
site which does not have a dimer touching it, and the classical
monomer-monomer correlator is the ratio of the number of
configurations with the two monomers to the number without
them. Monomers are deconfined if this ratio approaches a constant,
nonzero value as the distance between the two increases. In the
quantum dimer model, two monomers have the interpretation of two
``test spinons'' obtained by breaking a valence bond apart into its
constituent spins. Hence our result proves that spinons are deconfined
at high temperatures in the triangular lattice quantum dimer
model.\cite{fn-polyaloop,fn-werner} Second, we study the interpolation
between the critical square lattice and the non-critical triangular
lattice by tuning the fugacity for one species of bond. The diverging
correlation length near the former allows a continuum limit to be
taken and yields a theory of two Majorana fermions with the monomer
operator being identified with a linear combination of the spin
operators in the two sectors.

We begin by computing the partition function for general fugacities
and show that the model undergoes phase transitions only when it
reduces to the square lattice model. Correspondingly, it exhibits
long-ranged dimer correlations in the square lattice limit. Next we
derive asymptotic forms for the Green function and dimer correlations
which show that the correlation length is less than a lattice constant
at the isotropic point. We extend the analysis to obtain the
correlation length everywhere along the interpolation to the square
lattice. In Section IV we turn to computing the free energy of two
monomers in a background of dimers. We show that it falls off
exponentially to a constant, proving that spinons are deconfined in
the quantum dimer model at infinite temperature. In Section V we show
that the continuum formulation of the critical square lattice problem
is a theory of two Majorana fermions. The addition of the remaining
bonds of the triangular lattice is equivalent to adding a mass term
for each fermion and hence a transition of the Ising universality
class.  We end with some
concluding remarks and two technical appendices.

\section{Model and partition function}

A dimer is a bond connecting two nearest neighbors on a
lattice. We study the close-packed model with hard
cores, where an allowed dimer configuration has the property that each
site of the lattice is paired with exactly one of its nearest
neighbors, such a pair being denoted by a dimer placed on the link
between the two sites. In the simplest form of the model, each dimer
has the same fugacity; as the number of dimers is the same in all
configurations, the correlations of the dimer model are thus given as
the equally-weighted average over all possible dimer
configurations. In the following, we will include unequal fugacities,
so that the average to be taken then includes non-trivial weighting
factors.

The close-packed hard-core dimer model can solved on any planar
lattice by using Pfaffian techniques.  (A planar graph contains no
overlapping links.)  These techniques were introduced in the early
1960s by Kasteleyn,\cite{kasteleyn} and Fisher and
coworkers.\cite{fisherdimer,temperley,Fisher63}  The result is simple
to describe: on a planar graph one places arrows on the links so
that each plaquette is ``clockwise odd'', that is to say that the
product of the orientations of the arrows around any even-length
elementary plaquette traversed clockwise is odd. The antisymmetric
matrix $M_{ij}$ is then defined with $M_{ij}=1$ if an arrow points
from point $i$ to point $j$, $M_{ij}=-1$ if the arrow points from $j$
to $i$, and $M_{ij}=0$ if $i$ and $j$ are not nearest neighbors.
Kasteleyn's theorem states that for any planar graph $M$ can be
found, and that the number of dimer coverings (the partition function)
is given by the Pfaffian of the matrix $M$:
$$Z=\pm \hbox{Pf}[M].$$
The $\pm$ sign is chosen to make $Z$ positive; henceforth
we will omit this sign.
This result is exceptionally useful because the Pfaffian is the square
root of the determinant:
$$\hbox{Pf}[M] = (\det[M])^{1/2}.$$
For regular lattices, this determinant can easily be computed
by using Fourier transformation.
This result has been reformulated in the more modern language of
fermionic path integrals in Ref.~\onlinecite{samuel}. 
One places a Grassmann variable $\psi_i$ on each 
site $i$ of the lattice, and defines the action
$S=\sum_{i<j} M_{ij} \psi_i \psi_j$. Then a basic result of Grassmannian
integrals gives
$$Z=\int[{\cal D}\psi]\exp(S).$$ This form allows one easily to define
different but equivalent $M$ by rescaling the fermions.

\begin{figure}
\begin{center}
\leavevmode \epsfxsize 0.95\columnwidth \epsffile{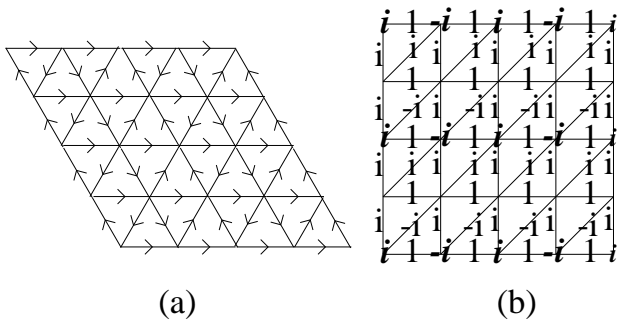}
\end{center}
\caption{Choice of $M$ for the triangular lattice. (a) A `clockwise
odd' sign convention. (b) By multiplying the Grassmann variables on the sites by
the phases indicated (italics), one obtains the new values for the
$M_{ij}$ on the bonds. Bonds are in addition multiplied by a
fugacity. Note that in both cases, the unit cell is
doubled.  }
\label{fig:signs}
\end{figure}

For the triangular lattice, an appropriate choice of arrows is
displayed in Fig.~\ref{fig:signs}.  To make contact with the work on
the square lattice, we however choose a different convention. First,
we deform the triangular lattice into a topologically (but not
symmetry) equivalent square lattice. Second, we multiply the Grassmann
variables on every other row by alternating factors of $\pm i$ so that
the values of $M_{ij}$\ are given as in Fig.~\ref{fig:signs}, where
the weight is to be understood to be for a convention of arrows
pointing to the right/upwards. Finally, in order to be able to
interpolate between square and triangular lattice, we allow the
fugacity of the diagonal bonds to be a variable $t$\ instead of 1. For
completeness, we give fugacities $v$\ and $u$\ to horizontal and
vertical bonds, respectively.  Note that we must double the unit cell
to contain two sites; in our convention, it is doubled in the vertical
direction.  Each site $i$ is labeled by the location of its unit
cell, $\bfx$, and its location in the basis, $\alpha$, while
the unit vectors between cells are $\hat\bfx$ and $\hat\bfy$. The two
fermions in the unit cell located at $\bfx$ are thus denoted as
$\psi_{\alpha,\bfx}$ with $\alpha=1,2$.  In this form, the action is \bea
S=\frac{1}{2}\sum_{\bfx,\alpha}\sum_{\bfy,\beta}
M^{\alpha\beta}_{\bfx\bfy} \psi_{\alpha,\bfx}\psi_{\beta,\bfy}
\label{eq:mat}
\eea
where $M^{\alpha\beta}_{\bfx\bfy}=-M^{\beta\alpha}_{\bfy\bfx}$.
With periodic boundary conditions (or far away from the boundaries), we have
$M^{\alpha\beta}_{\bfx\bfy}=M^{\alpha\beta}(\bfx-\bfy)$, where
\begin{eqnarray*}
M^{12}(0)&=&M^{21}(\hat\bfy)=iu,\\ 
M^{11}(\hat\bfx)&=&M^{22}(\hat\bfx)=v,\\
M^{21}(\hat\bfx+\hat\bfy)&=&-M^{12}(\hat\bfx)=it
\end{eqnarray*}

Since the action is quadratic in terms of the fermions, the model
can be solved by Fourier transformation. 
Our convention for
Fourier transforms is
$$
\widetilde{f}(\bfk)=\sum_{\bfx} \ e^{i\bfk\cdot \bfx} f(\bfx).
$$
The action is then
\bea
S= \frac12\sum_{\bfk,\alpha,\beta}\widetilde{M}_\bfk^{\alpha\beta}
\widetilde{\psi}_{\bfk,\alpha}\widetilde{\psi}_{-\bfk,\beta}\ ,
\eea
where the two-by-two matrix $\widetilde M_{\bf k}$ is
\begin{eqnarray*}
\widetilde M_{\bfk}=
\begin{pmatrix}
2iv \sin\kx& g(\bf k)\cr 
g^*(\bfk)&2iv\sin\kx
\label{eq:pmatrix}
\end{pmatrix}
\end{eqnarray*}
with\cite{fn-sign}
\begin{eqnarray*} 
g(\bfk) =i[u-t e^{i \kx}- ue^{-i \ky}-t e^{-i (\kx+\ky)}].
\end{eqnarray*}
Finding the determinant and hence the Pfaffian is now simple,
because in $\bfk$ space
the action is expressed in terms of the four-by-four blocks
\begin{equation}
\begin{pmatrix}
0&\widetilde M_{\bfk}\cr \widetilde M_{-\bfk}&0
\end{pmatrix}
\label{4by4}
\end{equation}
on the diagonal. Note that this matrix is antisymmetric as it must be,
because $g^*(\bfk)=-g(-\bfk)$.
The entropy per site, ${\cal S}$, of the dimer coverings on a
$N$-site lattice is then
\begin{equation}
{\cal S}=\frac1N \ln{\rm Pf}[M]=
\frac{1}{4}\sum_\bfk\ln\left|\Delta(\bfk)\right|,
\label{eq:det}
\end{equation}
where
\begin{eqnarray*}
\nonumber
\Delta(\bfk)&\equiv&\det[\widetilde M_\bfk] = \det[\widetilde M_{-\bfk}]\\
&=&
-4v^2 \sin^2\kx -4u^2\sin^2(\ky/2)\\
&&\qquad\qquad -4t^2\cos^2(\kx+\ky/2)
\end{eqnarray*}

\begin{table}
\begin{center}
\begin{tabular}{|c|l|c|c|}
\hline
$L_x$ & $L_y$ & $Z_{open}$ & $Z_{torus}$\\
\hline \hline
1 & 1 & 1 &  \\ 
2 & 1 & 2 &  \\ 
3 & 1 & 3 &  \\ 
4 & 1 & 5 & \\ 
5 & 1 & 8 &  \\ 
6 & 1 & 13 & \\ 
1 & 2 & 1 &  \\ 
2 & 2 & 5 &  \\ 
3 & 2 & 15 & 344 \\ 
4 & 2 & 56 & 1920 \\ 
5 & 2 & 203 & 10608 \\ 
6 & 2 & 749 & 59040 \\ 
1 & 3 & 1 &  \\ 
2 & 3 & 13 &  \\ 
3 & 3 & 85 & 4480 \\ 
4 & 3 & 749 & 59040 \\ 
5 & 3 & 6475 & 767776 \\ 
1 & 4 & 1 &  \\ 
2 & 4 & 34 &  \\ 
3 & 4 & 493 & 58592 \\ 
4 & 4 & 10293 & 1826944  \\ 
1 & 5 & 1 & \\ 
2 & 5 & 89 & \\ 
3 & 5 & 2871 & 766528 \\ 
1 & 6 & 1 &  \\ 
2 & 6 & 233 & \\ 
3 & 6 & 16731 &  10028288   \\ 
\hline
\end {tabular}
\end{center}
\caption{Number of dimer coverings, $Z$, of triangular lattices of
with $L_x\times\ 2 L_y$\ sites, with periodic and open boundary
conditions. For periodic boundary conditions, only those sizes 
$(L_x\geq3\ {\rm and}\ L_y\geq 2)$\ are given in which any pair of
sites is linked by at most one bond.
}
\label{tab:dimcov}
\end{table}

For a system with periodic boundary conditions, one must sum
over four different sectors, according to the winding of the dimer
configuration.\cite{kasteleyn} In fermionic language, this corresponds to
evaluating four Pfaffians, with symmetric and antisymmetric boundary
conditions, and combining them as $Z=(-{\rm Pf}_{pp}+{\rm
Pf}_{ap}+{\rm Pf}_{pa}+{\rm Pf}_{aa})/2$,
where the subscripts denote the
(anti)periodic boundary conditions for the fermions in the two
directions. In a Fourier representation, this
corresponds to the usual different choices of allowed wavevectors.  We
have listed the number of dimer coverings thus obtained for the
isotropic case in Table~\ref{tab:dimcov}, along with the results for a
system with open boundary conditions.

In the thermodynamic limit, the sum over $\bfk$ in (\ref{eq:det})
turns into an integral.
For the isotropic triangular lattice ($t=u=v=1$), 
doing the integral numerically yields
$${\cal S} = 0.4286\dots .$$
We note ${\cal S}$ has been obtained previously by several
authors\cite{nagle,gaunt,samuel,wolffzitt,mila} in different contexts,
including the kagome Heisenberg magnet\cite{mila} and, implicitly,
the fully frustrated Ising model on the hexagonal lattice,\cite{wolffzitt} 
where the ground state entropy per spin ${\cal S}_{FFHIM}={\cal S}/2$ 
is related to the dimer model entropy via a
duality mapping.\cite{duality}

This model simplifies in several limiting cases. If two fugacities
vanish, the model reduces to decoupled chains, with order along the
chains and disorder relative to one another. More interesting is the
case where one of the fugacities vanishes, the square-lattice dimer
model.  This model is critical, with algebraic decay
of correlators.\cite{Fisher63,Hartwig} It is straightforward to
show that the entropy
${\cal S}(t,v,u)$ is non-analytic in $t$ at $t=0$ (and likewise for $u$
and $v$). We will discuss this in more detail in section V.

\section{The Green function}

We now turn our attention to the correlations of the triangular dimer
model. We utilize the techniques developed for the square lattice in
Ref.~\onlinecite{Fisher63}, and expressed in terms of Grassmann variables in 
Ref.~\onlinecite{samuel}.
Correlation functions can
be straightforwardly expressed in terms of the Grassmann variables
(see e.g.\ Ref.~\onlinecite{slowhole}).
Since the action is quadratic, using Wick's theorem
expresses all correlators in terms of the Green function
$$\langle \psi_i \psi_j \rangle
=\frac{1}{Z}\int[{\cal D}\psi]\ \psi_i\psi_j\exp(S).$$ 
For example, the probability $P^{(d)}(\hat{\bf j})$
of finding a dimer on a bond in the $(\hat{\bf j})$ direction at site $0$
is given by
\begin{equation}
P^{(d)} (\hat{\bf j}) = \big| \langle \psi_{0} \psi_{\hat {\bf j}}
\rangle\big| \ .
\label{pd1}
\end{equation}
In the original matrix language,
the Green function elements make up the inverse of the matrix $M$,
namely
$$
\langle \psi_i \psi_j\rangle = (M^{-1})_{ji} = - (M^{-1})_{ij}.
\label{green}
$$
In this section we explicitly determine the Green function
and its asymptotic behavior. In the next we use these results
to determine the dimer-dimer and monomer-monomer correlators.

As shown in the last section, in Fourier space
the action is written in terms of the four-by-four blocks (\ref{4by4}). 
These blocks can be easily inverted to give the two-point functions
in Fourier space.
Expressed in terms of the functions $g(\bfk)$ and $\Delta(\bfk)$,
we have
\begin{eqnarray*}
\langle \widetilde\psi_{1,\bfk}\widetilde\psi_{1,-{\bf k}}\rangle &=& 
\langle \widetilde\psi_{2,\bfk}\widetilde\psi_{2,-{\bf k}}\rangle = 
\frac{2i \sin(k_x)}{\Delta(\bfk)}\\
\langle \widetilde\psi_{1,\bfk}\widetilde\psi_{2,-{\bf k}}
\rangle &=& -\frac{g^*(\bfk)}{\Delta(\bfk)}\\
\langle \widetilde\psi_{2,\bfk}\widetilde\psi_{1,-{\bf k}}\rangle 
&=& -\frac{g(\bfk)}{\Delta(\bfk)}
\end{eqnarray*}
To determine the Green functions in real space, we need to invert
the Fourier transform. This yields
\begin{equation}
 \langle \psi_{\alpha,\bfx}\psi_{\beta,{\bf \bfy}}\rangle =
\int d\bfk\ e^{-i\bfk\cdot(\bfx-\bfy)}
\langle \widetilde\psi_{\alpha,\bfk}\widetilde\psi_{\beta,-{\bf k}}\rangle
\end{equation}
where we define
$$\int d{\bf k} \equiv \frac{1}{4\pi^2}
\int_0^{2\pi} dk_x \int_0^{2\pi} dk_y .$$
For a finite number of sites, this integral is replaced with the usual sum.

To simplify the resulting expressions, we specialize slightly in the
subsequent analysis:
\begin{enumerate}
\item We work in the thermodynamic limit so we have integrals over $\bfk$.
\item We set $u=v=1$. Thus the
fugacity $t$ gives a way of interpolating between the square and the
isotropic triangular lattices, with $t=0$ giving the former and $t=1$ the
latter.
\item We study correlators where the fermions are in the same or
adjacent rows. 
\end{enumerate}
It is also convenient to absorb some factors of $\pm i$ by defining the 
Green functions $Q_s$ and $R_s$ as
\begin{eqnarray*}
\langle \psi_{1,\bfx} \psi_{1,\bfx+s\hat\bfx} \rangle &=&
\langle \psi_{2,\bfx} \psi_{2,\bfx+s\hat\bfx} \rangle 
\equiv (-1)^{(s+1)/2} Q_s \\
\langle \psi_{1,\bfx} \psi_{2,\bfx+s\hat\bfx}\rangle &=&
-\langle \psi_{2,\bfx} \psi_{1,\bfx-s\hat\bfx}\rangle  
\equiv i(-1)^{[(s+1)/2]} R_s
\end{eqnarray*}
where $[a]$ is the greatest integer less than $a$.
Relatively simple expressions for $Q_s$ and $R_s$ are obtained
by shifting $k_x\to k_x +\pi/2$ and   $k_y\to 2k_y +\pi$. 
For odd $s$
\begin{eqnarray} 
\label{Qodd}
Q_s &=& \int d{\bf k}\ w(k_x,k_y) \cos(k_x) \cos(sk_x)\\
\label{Rodd}
R_s &=& t\int d{\bf k}\ w(k_x,k_y) \cos(k_x+k_y) \cos(sk_x+k_y)
\end{eqnarray}
while for even $s$
\begin{eqnarray} 
Q_s &=& 0 \\
\label{Reven}
R_s &=& \int d{\bf k}\ w(k_x,k_y) \cos(k_y) \cos(sk_x+k_y) 
\end{eqnarray}
where
$$w(k_x,k_y) = \frac{1}{2}\frac{1}{\cos^2(k_x) + \cos^2(k_y) 
+ t^2\cos^2(k_x+k_y)}.
$$
These integrands are all invariant under the interchange of $k_x$ and
$k_y$, under ${\bf k} \to -{\bf k}$, and under shifts of $\pi$.

One of the two integrals in each of the
$Q_s$ and $R_s$ can be done immediately by residue.
We let $a=k_x+k_y$ and $b=k_x-k_y$ so that
$w^{-1}=2(1+\cos(a)\cos(b)+t^2\cos^2(a))$. 
The integral over $b$ can be deformed so that the contour runs
around the pole at 
$$\cos(b)=\frac{1+t^2\cos^2(a)}{\cos(a)}.
$$
This is done by changing the integration contour running from $-\pi$\ to
$\pi$ along the real axis to one which first runs from $-\pi$\ to
$-\pi+i\infty$, then to $+\pi+i\infty$, and finally back down to
$\pi$. The two contours running parallel to the imaginary axis cancel
as the function is periodic under $b\rightarrow b +2\pi$, 
and the contribution from the contour running parallel to the real axis
vanishes when the integrand vanishes exponentially as Im$(b)\to\infty$.
We thus pick up only the contribution of the pole of the integrand.
For example, we have
$$\int_0^{2\pi} \frac{db}{2\pi} \frac{1}{1+r\cos(b)+t^2r^2} = 
\frac{1}{\sqrt{1+(2t^2-1)r^2+t^4r^4}}.$$

The resulting expressions for $Q_s$ and $R_s$ are simpler on the
isotropic triangular lattice $t=1$ because of the additional symmetry.
In particular, here $w(k_x,k_y)$
is invariant under the transformation $k_x\to -(k_x+k_y)$; this
ensures that the Green functions are invariant under 60-degree
rotations of the lattice. For example, this transformation allows us
to change the numerator of $Q_s$ from $\cos(sk_x)\cos(k_x)$ to
$\cos(sa)\cos(a)$. This gives us
\begin{equation}
Q_s (t=1) = \frac{1}{2\pi}\int_0^\pi  da \frac{\cos(sa)\cos(a)}
{\sqrt{1+\cos^2(a)+\cos^4(a)}}
\label{qodd}
\end{equation}
and for odd $s$
\begin{equation}
R_s (t=1) = \frac{1}{4\pi}\int_0^\pi  da \frac{\cos(sa)e^{ib_0}+\cos((s-1)a)}
{\sqrt{1+\cos^2(a)+\cos^4(a)}}
\label{rodd}
\end{equation}
while for even $s$
\begin{equation}
R_s (t=1) = \frac{1}{4\pi}\int_0^\pi  da \frac{\cos((s-1)a)e^{ib_0}+\cos(sa)}
{\sqrt{1+\cos^2(a)+\cos^4(a)}}
\label{reven}
\end{equation}
where
$$e^{ib_0}=\frac{1+\cos^2(a)-\sqrt{1+\cos^2(a)+\cos^4(a)}}{\cos(a)}.$$

\subsection{Asymptotic long-distance behavior}
\label{subsect:longdist}

In this subsection, we study the long-distance behavior
of the Green functions $Q_s$ and $R_s$. We derive an expression
for the correlation length of the fermions
at any $t$, and find that it indeed vanishes only when $t=0$.

\begin{figure}
\begin{center}
\leavevmode \epsfxsize 0.9\columnwidth \epsffile{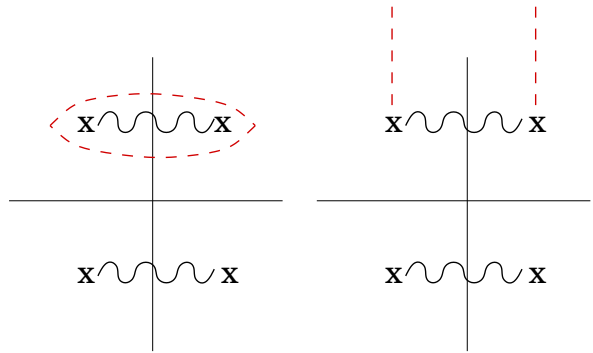}
\end{center}
\caption{The integration contours (dashed) used for the asymptotic
evaluation of the Green function when $t>1/2$. The second is equivalent to
the first if the integrand is multiplied by 2. For $t<1/2$, the 
branch cuts have Re$(a)=0$.}
\label{fig:contour}
\end{figure}

Unlike the similar case of the square lattice, the integrals 
for the Green function cannot be
evaluated asymptotically by partial integration. This method only
works for algebraic correlations, whereas in our case, the
correlations decay exponentially. One can nonetheless obtain the
asymptotics by deforming the integration contour. Instead of there
being poles in the complex $a$-plane, there are now square-root
branch cuts. In the region $|\hbox{Re}(a)|<\pi/2$, there is
one branch cut in the lower-half-plane and one in the upper.
For $t<1/2$, the branch cut has Re$(a)=0$, while for $t>1/2$, it
is parallel to the Re$(a)$ axis. The general formula for the
location of the branch points $a_{\pm}$ is
\begin{equation}
\cos(a_\pm) = \frac{1\pm\sqrt{1-4t^2}}{2t^2}
\label{branch}
\end{equation}
For the same reason we could deform the $b$-integral around the
pole, we can now deform the $a$-integral around the branch cut,
as depicted in first part of Fig.~\ref{fig:contour}. To evaluate
the integral by saddle point, we want a contour where the $e^{isa}$
term in the integrand is decreasing exponentially. It is thus
easiest to use the fact that with a square-root branch cut, the
integral around one side of the branch is equal to the integral around
the other. Thus the contour can be deformed yet again, so that it
goes from the two branch points to $i\infty$, as long as we multiply
the result by 2. This contour, denoted by ${\cal C}$,
is depicted in the second part of
Fig.~\ref{fig:contour}.

With this new contour, asymptotic evaluation of the integral is easy,
because the integrands are sharply peaked at the branch points. 
For example, we have
$$Q_s(t=1)= \frac{1}{2\pi} \int_{\cal C} da
\frac{e^{i(s+1)a}+e^{i(s-1)a}} {\sqrt{1+\cos^2(a)+\cos^4(a)}}.$$ The
exponentials rapidly decrease as $\hbox{Im}(a)$ is increased along the
contour, and moreover, the rest of the integrand has a square-root
divergence at the branch points. It is thus an asymptotically exact
approximation to substitute ${\sqrt{1+\cos^2(a)+\cos^4(a)}}\approx
C\sqrt{a-a_\pm}$ into the integrand.  This gives
\begin{equation}
Q_s(t=1)\approx f(|s+1|)+f(|s-1|)
\label{asymq}
\end{equation}
where
$$f(s)\equiv
\hbox{Re }\left[ \frac{1}{\sqrt{C}\pi}
\int_{a_+}^{i\infty} da \frac{e^{-isa}}{\sqrt{a-a_+}}\right].$$
We have used the fact that the integral from $a_-$ to
infinity is the complex conjugate of the integral from $a_+$ to
infinity. After shifting $a\to a+a_+$, the integral
is easily done, yielding
$$f(s)= \hbox{Re }\left[\frac{1}{\sqrt{Cs\pi}}e^{ia_+s}\right].$$
Finally, we break $a_+$ into its real and imaginary parts:
$a_+ = \alpha+i\beta$. For $t=1$, $\cos(a_+) = \exp(2i\pi/3)$,
so that $C=3^{3/4}2\exp({-i\pi/4})$, $\alpha=1.1960\dots$ and
\begin{equation} 
\beta= \frac{1}{4}\ln\left[\frac{\sqrt{2}+3^{1/4}}{\sqrt{2}-3^{1/4}}\right]
=.83144\dots
\end{equation}
Plugging this in gives
\begin{equation}
f(s)= \frac{1}{\sqrt{2\pi 3^{3/4}}} \frac{e^{-\beta s}}{\sqrt{s}}
 \cos\left(\alpha s+ \frac{\pi}{8}\right) 
\label{asym}
\end{equation}
The asymptotic expression for $R_s$ can now easily be found by noticing
that at the branch point $a=a_+$, $e^{ib_0}=-1$. This means that
\begin{equation}
R_s (t=1)\approx (-1)^s \big[f(|s|)-f(|s-1|)\big]
\label{asymr}
\end{equation}

\begin{figure}
\begin{center}
\leavevmode \epsfxsize 0.9\columnwidth \epsffile{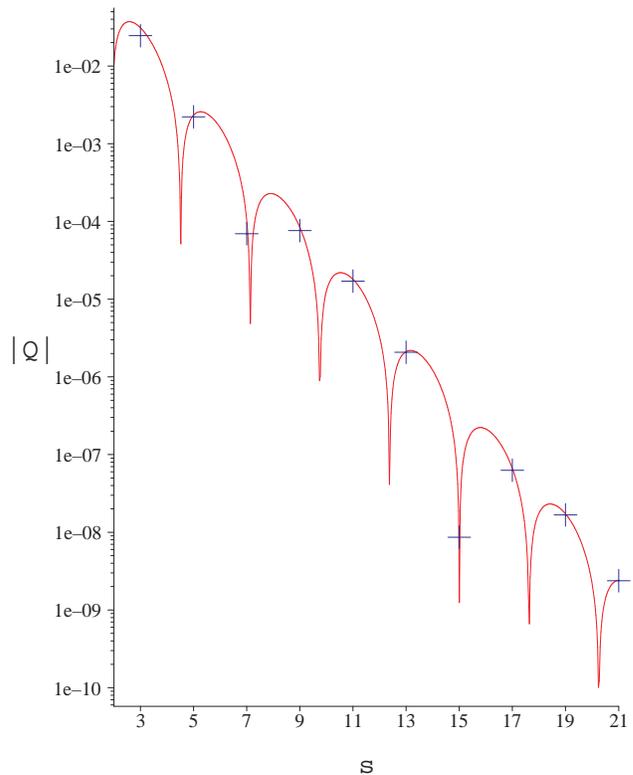}
\end{center}
\caption{The curve is the asymptotic expression for $|Q_s|$ when $t=1$
(\ref{asymq}).
The points are the exact values at odd $s$; $Q_s$ vanishes
for even $s$.}
\label{fig:intcomp}
\end{figure}

We have therefore shown that on the isotropic triangular lattice,
$\beta$ is the inverse correlation length along the rows. It is only
about one lattice spacing, so the exponential decay of the Green
functions is quite rapid.  Comparing the asymptotic form to the exact
values derived in the next subsection (see Fig.~\ref{fig:intcomp} for
a plot of $Q_s$), one finds that the agreement is excellent---within a
few percent---even for small values of $s$. At occasional values of $s$
(e.g.\ $s=15$), however, the agreement is only within a factor of $2$
or so. The reason is that the oscillating factor in (\ref{asym})
occasionally becomes very close to zero near $s$ an odd integer, so
that the terms we have neglected above can become larger than the one
we kept. This however, does not change our results for the correlation
length, because the neglected terms have the same exponential
dependence on $s$ (but a power law exponent different from $-1/2$ in
Eq.~\ref{asym}).

\begin{figure}
\begin{center}
\leavevmode \epsfxsize 0.9\columnwidth \epsffile{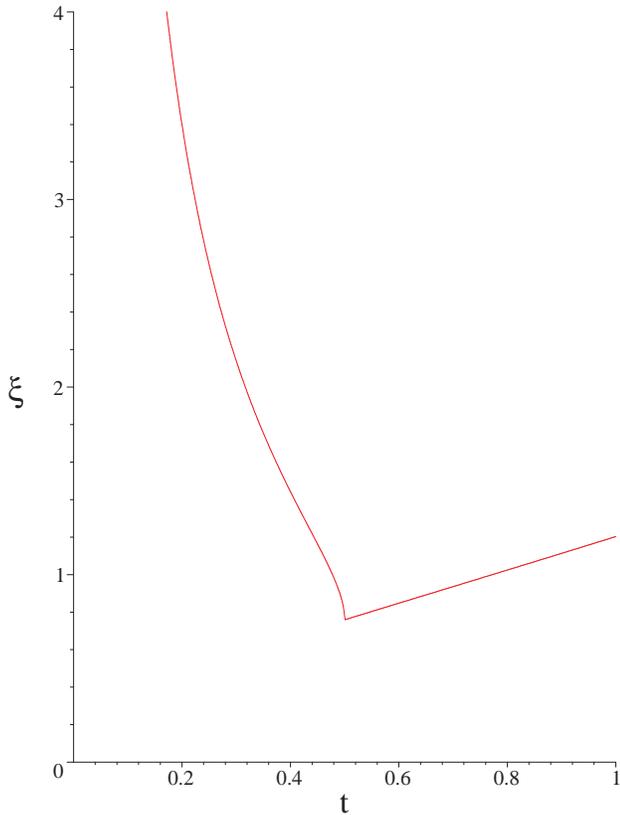}
\end{center}
\caption{ 
The correlation length as a function of the dimer fugacity:
$t=0$ is the square lattice, $t=1$ the isotropic triangular lattice.
}
\label{fig:corr}
\end{figure}

For arbitrary $t$, the computations are similar but the equations look
fairly gruesome. Finding asymptotic expressions for $Q_s$ and $R_s$ is
more complicated because the transformation $k_x \to -(k_x+k_y)$ no
longer leaves $w(k_x,k_y)$ invariant.  However, we can easily extract
the correlation length for Green functions in the $\hat \bfx
+\hat\bfy$ direction (the direction along the links with $t$-dependent
fugacity).  The reason is that the expressions for the Green
functions in this direction end up very similar to
(\ref{rodd},\ref{reven}).  The key fact is that there are no terms in
the integrand involving $e^{isb_0}$, only $e^{ib_0}$. Like before,
the integrand is peaked around the branch points given in
(\ref{branch}). Thus the fermion correlation length 
$\xi^{\hat \bfx +\hat\bfy}$ is given by
\begin{equation}
(\xi^{\hat \bfx +\hat\bfy})^{-1}=\hbox{Im}\left[
\arccos\left(\frac{1-\sqrt{1-4t^2}}{2t^2}\right)\right].
\label{corr}
\end{equation}
We plot this correlation length in Fig.\ \ref{fig:corr}.
Note that the kink at $t=1/2$ in Fig.\ \ref{fig:corr} is evidence for
a level crossing.

As $t\to 0$, the correlation length diverges as $1/t$:
$$\xi^{\hat \bfx +\hat\bfy} \approx \frac{1}{\sqrt{2} t}\qquad\qquad
t\to 0$$ while $\xi=1/\beta=1.2027\dots$ at $t=1$. In this expression,
we have taken the length of a diagonal bond as unit
distance. Translating this back into square-lattice language (with the
length unit given by a lattice constant) removes the factor of
$1/\sqrt{2}$.  The diverging correlation length in the square lattice
limit will be used to take a continuum limit in Section V below.

\subsection{Short-distance behavior on the isotropic lattice}
\label{subsect:shortdist}

In this subsection we show how to derive exact expressions for the
Green functions on the isotropic triangular lattice, $t=1$.  We
start by noting that a few of these Green functions can be evaluated
immediately. In the isotropic case, 
the probability of a dimer being on a given
link is $1/6$. This is of course related to the Green
function for neighboring fermions by (\ref{pd1}) so
$$Q_1 = R_1 = R_0 = \frac{1}{6}.$$
Ironically, this fact is not easy to extract from the explicit
integrals (\ref{qodd},\ref{rodd},\ref{reven}). It does follow from
noting that the transformation $k_x \to -(k_x+k_y)$
in (\ref{Qodd},\ref{Rodd},\ref{Reven}) means that
$Q_1=R_1=R_0$, and also that $Q_1+R_1+R_0=1/2$.

By generalizations of this argument, 
it is possible to derive
recursion relations relating $Q_s$ and $R_s$. For
the square lattice they are discussed in
Ref.~\onlinecite{Hartwig}.  For the isotropic triangular lattice, the
relations relating all the $R_s$ to the $Q_s$ can be derived using
simple trigonometric identities.  One finds here that
$$R_{2j+2}=R_{2j+1} - R_{2j} + R_{2j-1}+ Q_{2j+1} - Q_{2j-1}$$
and
$$2R_{2j+1} = -2R_{2j} - Q_{2j+1} - Q_{2j-1} + \delta_{j0}.$$ 
The Kronecker delta in the latter identity arises from the integral $\int
d{\bf k}\ \cos(jx) =\delta_{j0}$.  The recursion relations involving
only $Q_s$ are trickier to obtain; we discuss them in the Appendix.

The recursion relations mean that $Q_s$ and $R_s$ for any $s$ can be
expressed in terms of $Q_1$, $R_2$ and $Q_3$.  These in turn can be
evaluated in terms of elementary functions; in particular the
combination
\begin{equation}
U_0 = \frac{\Gamma(7/6)}{2\Gamma(2/3)\sqrt{\pi}}
\end{equation}
As detailed in Appendix A, along with $Q_1=1/6$ we have
\begin{eqnarray*}
R_2&=&\frac{1}{3}-2U_0\\
Q_3&=&-\frac{1}{2}+\frac{1}{6}\,{\frac {\sqrt {3}}{{\it U}_{{0}}\pi }}
\end{eqnarray*}
Using these facts, we can evaluate the Green functions to
arbitrarily good accuracy numerically by using e.g.\ Maple. Although
it is easy to iterate the recursion relations on the computer, we have
not succeeded in finding a closed-form expression for arbitrary
$Q_s$ and $R_s$.  We have collected some of them in Table
\ref{tab:Gs}. Note that even though the coefficients are increasing
exponentially with $s$,
the three terms in each cancel almost perfectly to give the
exponential falloff in $s$. 

\begin{table}
\begin{center}
\begin{tabular}{|c|c|c|}
\hline
$s$ & $Q_s$ & numerical value\\
\hline\hline
$1$&$
1/6$&$
 0.166666666666666666667$\\$
3$&$
-1/2+1/6\,{\frac {\sqrt {3}}{{\it U}_{{0}}\pi }}$&$
- 0.024550581458226763328$\\$
5$&$
8\,{\it U}_{{0}}+5/6-5/6\,{\frac {\sqrt {3}}{{\it U}_{{0}}\pi }}$&$
 0.0022132632430383043346$\\$
7$&$
-56\,{\it U}_{{0}}+{\frac {25}{6}}+7/3\,{\frac {\sqrt {3}}{{\it U}_{
{0}}\pi }}$&$
 0.0000693679214938995513$\\$
9$&$
216\,{\it U}_{{0}}-{\frac {93}{2}}+5/3\,{\frac {\sqrt {3}}{{\it U}_{
{0}}\pi }}$&$
- 0.000076203880846465563$\\$
11$&$
-{\frac {1288}{5}}\,{\it U}_{{0}}+{\frac {1397}{6}}-{\frac {385}{6}}
\,{\frac {\sqrt {3}}{{\it U}_{{0}}\pi }}$&$
 0.0000170664326460443169$\\$
13$&$
-{\frac {16016}{5}}\,{\it U}_{{0}}-{\frac {3443}{6}}+{\frac {2509}{6}
}\,{\frac {\sqrt {3}}{{\it U}_{{0}}\pi }}$&$
- 0.000002068500172729645$\\$
15$&$
29232\,{\it U}_{{0}}-{\frac {2885}{2}}-{\frac {30970}{21}}\,{\frac {
\sqrt {3}}{{\it U}_{{0}}\pi }}$&$
 0.0000000086255295387778$\\
\hline
\end {tabular}
\end{center}
\caption{The Green function $Q_s$ explicitly}
\label{tab:Gs}
\end{table}

\section{Correlators}

\subsection{The dimer-dimer correlation function}

The dimer-dimer correlation function is easily expressed in terms of
the Green functions. The operator $\psi_{1,\bfx}\psi_{2,\bfx}$
creates a dimer at site $\bfx$ pointing in the $\hat\bfy$ direction. 
The probability that there are two parallel dimers in the same row a distance
$s$ apart is therefore
$$P^{(dd)}(s)=\langle \psi_{1,\bfx}\psi_{2,\bfx} 
\psi_{1,\bfx+s\hat\bfx}\psi_{2,\bfx +s\hat\bfx}\rangle$$
By Wick's theorem, this decomposes into
$$P^{(dd)}(s)=(R_0)^2 -(Q_s)^2 - R_s R_{-s}$$
This therefore decays exponentially to its asymptotic value with half
the correlation length of the Green function.
For the isotropic lattice, one can easily plug in the asymptotic forms 
(\ref{asymq},\ref{asymr}) derived in the last section, and find the
correlation length $1/(2\beta)\approx 0.6014$, less than one lattice
spacing.

\subsection{The monomer-monomer correlation function}

As noted in the introduction, the asymptotic behavior
of the monomer-monomer correlation
function is quite interesting physically. A zero value means
the model is confining, while a nonzero value means that the model
is in a deconfined phase. We will show in this subsection that
on the isotropic triangular lattice, the latter is true.

\begin{figure}
\begin{center}
\leavevmode \epsfxsize .94\columnwidth \epsffile{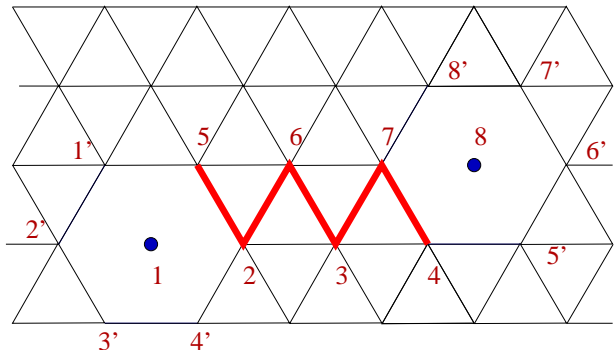}
\end{center}
\caption{The string in the monomer-monomer correlator with $p=4$. 
}
\label{fig:mono}
\end{figure}

A monomer placed on a lattice site forbids a dimer from being placed
on any of the links connected to the site.  The monomer-monomer
correlator $P^{(mm)}(q,r)$ is the ratio of the number of
configurations with monomers at sites $q$ and $r$ to the number of
configurations without the monomers.  Thus computing a monomer-monomer
correlator follows from the partition function of the lattice with the
two sites (and all the links connected to these sites) deleted.  Since
such a lattice is planar, Kasteleyn's construction is still
applicable.  The one complication is that we must ensure that on the
lattice with deleted sites, the number of arrows is still clockwise
odd.  With the assignment in Figure 1, one sees immediately that the
number of arrows around a deleted site is even. Thus this assignment
must be modified by reversing one of the arrows in the plaquette
around the deleted site. Reversing the arrow on this link then ruins
the clockwise-odd assignment around the other plaquette this link
borders. Thus we must reverse one of the other arrows on this other
plaquette. This in turn ruins another assignment, and so on.  We thus
must build up a string of reversed arrows, which must stretch from one
monomer to the other. As long as the arrows are chosen to make
all plaquettes clockwise odd, the monomer correlator is
independent of the choice of path of the string.
For monomers in adjacent rows, this construction
is illustrated in Fig.\ \ref{fig:mono}; links with the thick lines are
part of the string.

This is very much like constructing the spin-spin correlation function
in the Ising model in terms of fermionic variables. In fact, on the
square lattice, the monomer-monomer correlator was shown
\cite{Fisher63,Hartwig} to have the same long-distance behavior
(falloff as the square root of the distance) as the spin-spin
correlation function in two decoupled Ising models. The precise
relation between the two correlators for the square lattice was given
in Ref.\ \onlinecite{AYP}.  In section V, we will extend
this relation away from the square lattice.

We define as the monomer-monomer correlation function
$P^{(mm)}$ as the ratio of the number of configurations
with the monomers to the number without.
This can be written in terms of the fermions.
Deleting a site merely corresponds to inserting a fermion at that site.
Changing the sign of the arrow corresponds to changing the sign of the 
term of the fermion in the action. In our theory with action
$\sum_{j<k} M_{jk} \psi_{j}\psi_k$, for a monomers at sites $q$ and $r$, 
\begin{equation}
P^{(mm)}(q,r) = \big|\langle \psi_q \prod(1-2M_{ab}\psi_a\psi_b) \psi_r
 \rangle\big|
\end{equation}
where the product is over all the links connecting sites
$q$ and $r$ which have reversed arrows.  We specialize to
the case of one monomer being at the origin and the other in an
adjacent row $p$\ lattice spacings apart, as illustrated
in Fig.\ \ref{fig:mono}. Labeling the fermions in the lower row
$1\dots p$ and those in the upper row as $p+1\dots 2p$ gives
\begin{eqnarray}
\nonumber
P^{(mm)}(p) &=& \big|\langle \psi_{1} 
(1-2i\psi_{p+1}\psi_{2})(1+2it\psi_{2}\psi_{p+2})\\
&&\qquad \dots(1-2i\psi_{2p-1}\psi_{p})\,
\psi_{2p}\rangle\big|
\label{mmdef}
\end{eqnarray}
Since the theory is free in terms of the fermions, one can use Wick's
theorem to express this correlator as a product of Green
functions. Evaluating it by brute force, however, is too difficult for
all but low values of $p$, because the number of contractions grows
exponentially. For a two-site separation,
$$P^{(mm)}(2)= R_2(1-R_0)-2R_1^2-2Q_1^2.$$
Using the results of the previous section for the Green functions, one has 
for $t=1$
$$P^{(mm)}(2)= 4U_0/3 - 1/9 = .14657672599\dots$$

To proceed further, one must in revert to the Pfaffian methods
developed for the square lattice.\cite{Fisher63} Our
calculation generalizes these methods to the triangular lattice.
The monomer-monomer correlator (\ref{mmdef}) can be expressed in
Pfaffian language as
$$P^{(mm)}=\frac{\hbox{Pf}(M^{(mm)})}{\hbox{Pf}( M)}$$ where
$M^{(mm)}$ is the Kasteleyn matrix with the monomer sites removed, and
the signs reversed on the links of the string connecting the monomers.
The advantage of this formulation is that one can manipulate
a matrix without changing the determinant. For example, if one
adds any row of a matrix to any other row, the determinant is left
unchanged. Thus the strategy is to find explicitly
the matrix $M^{-1}M^{(mm)}$ describing the monomer correlator, and
manipulate it to make it more tractable. \cite{Fisher63}  
On the square lattice, it is simple to show that
$$\det(M^{-1}M^{(mm)}) = \det(T)^2,$$ where 
$T$ is a $p\times p$ matrix.  After a variety of manipulations
described in appendix B, one finds the same behavior on the
triangular lattice.

Deferring the details to appendix B, our result is
that the monomer-monomer correlator is
\begin{equation}
P^{(mm)}(p) = \frac{1}{2} \det({\cal R}+{\cal Q}) .
\label{eq:monodet}
\end{equation}
where the entries of the $p\times p$ matrices ${\cal Q}$ and ${\cal R}$ are
\begin{eqnarray}
{\cal R}_{ij}&=&(-1)^{[(j-i)/2]}R_{j-i+1} + t^{i-j-1}\theta(i-j)\\
{\cal Q}_{ij}&=&i(-1)^{[(j+i)/2]}Q_{p+1-i-j} 
\end{eqnarray}
where $\theta(x)=1$ for $x>0$ and
$0$ for $x\le0$. 
Unlike the case of the square lattice, the matrix ${\cal R}+{\cal Q}$
is not Toeplitz (the entries in a Toeplitz matrix $T_{ij}$ depend only
on $i-j$).  For Toeplitz matrices, Szego's theorem provides a simple
way of finding the asymptotic behavior of the determinant as the
matrices get large. The theorem is not applicable here, because
while ${\cal R}_{ij}$ depends only on $i-j$, 
${\cal Q}_{ij}$ depends on $i+j$.

\begin{table}
\begin{center}
\begin{tabular}{|c|c|}
\hline
$s$ & $P^{(mm)}(s)$ \\
\hline\hline
1&
 0.166666666666666666666667\\
2&
 0.146576725991984081282220\\
3&
 0.150263558036976118604604\\
4&
 0.149354528957809826733084\\
5&
 0.149441157286260650652237\\
6&
 0.149430105031348455969637\\
7&
 0.149429091105386654359851\\
8&
 0.149429312376742966925414\\
9&
 0.149429243123620452470446\\
10&
 0.149429245483558265741555\\
11&
 0.149429245691495770851350\\
12&
 0.149429245319212097747918\\
13&
 0.149429245371125328892038\\
14&
 0.149429245361581959940523\\
15&
 0.149429245361256645973930\\
\hline
\end{tabular}
\end{center}
\caption{The monomer-monomer correlation function
for the isotropic triangular lattice}
\label{tab:mm}
\end{table}

To evaluate this correlator for $t=1$, 
we can plug the Green functions derived
in the previous section into the expression for $P^{(mm)}$
(\ref{eq:monodet}). Expressing them in terms of $U_0$ as before, the
final form of the determinant simplifies substantially (although it is
still fairly horrible). For example, for $p=3$ one finds
$$P^{(mm)}(3)=\frac{\sqrt{3}+(2\pi-3\sqrt{3})U_0
-144\pi U_0^3 + 864\pi U_0^4}{27\pi U_0}.$$
We have collected some numerical values in Table \ref{tab:mm}.
The correlator falls off exponentially to the value
$$P^{(mm)}(\infty)= 0.14942924536134225401731517482693\dots$$ 
To all $30+$ digits, this number is given by \cite{isc}
$$P^{(mm)}(\infty)= \frac{\sin(\pi/12)}{\sqrt{3}}.$$ 
Since this number
is fairly simple (and involves $\sin(\pi/12)$, which has already
appeared in our analysis), we expect that this expression
is exact and there exists
a generalization of Szego's theorem applicable to the determinant
(\ref{eq:monodet}).

The non-vanishing of the monomer
one-point function ($\equiv\sqrt{P^{(mm)}(\infty)}$)
means that the monomers are deconfined in this phase.
Numerically, one can check that the correlation length here is the
same as for the dimer-dimer correlation, namely $1/(2\beta)$, or about
$.6$ of a lattice spacing.

\subsection{The vison-vison correlation function}

In the quantum dimer model, a $\pi$-flux vortex or 
``vison''\cite{ReadCha,Kivelson} involves a semi-infinite
string on the dual lattice. A particular dimer configuration $c$
acquires a factor of $(-1)^{N_{\rm string}(c)}$, where 
$N_{\rm string}(c)$ is the number of dimers intersecting that 
string. For two visons separated by ${\bf x}$ 
the string can be taken to run from one
vison to the other and hence no flux is visible at infinity.
At the RK point, the calculation of the static two vison 
correlator reduces to the evaluation of 
$\sum_c (-1)^{N_{{\rm string}(0, {\bf x})}(c)}$ 
which is now a purely classical problem. We will henceforth refer to 
this sum and its generalizations to unequal fugacities as the two-vison 
correlator even though the latter do not correspond to any known quantum
problem. We should also note that the choice of string, which needs
to be made for each separation, lead to an ambiguity of a sign at each
separation.

The two-vison correlator
was studied recently in Ref.~\onlinecite{bfg} for a Kagome spin problem 
equivalent to a generalized dimer model on the triangular lattice
requiring three dimers per site. There are some special features of this
problem, arising from the ``particle-hole'' symmetry of occupying three
dimers out of a possible six at each site and among them is the presence
of two species of visons (for details the reader should consult the 
original work). In our problem, that of one dimer per site on the
triangular lattice, life is simpler. Here the two-vison correlator can be 
seen, most simply by duality arguments described in Appendix A of 
Ref.~\onlinecite{duality}, to be equal in magnitude (we have already
commented on the sign ambiguity) to the
spin-spin correlation function averaged over the ground state manifold
of the fully frustrated Ising model on the honeycomb lattice. The latter
is known, and decays exponentially with a correlation length computed
in Ref.\ \cite{wolffzitt}.
A direct computation of the vison correlator
is in progress \cite{shtengel} and should obviate the need for
the detour we have taken in this section.

\section{Perturbing about the square lattice: field theory}

As we noted earlier, the general triangular lattice dimer model is
critical in the square lattice limit. This suggests
that we can gain some insight into its properties by considering a
continuum limit when the diagonal fugacity $t$ is small. We should
immediately enter one caveat: our exact computation shows that there
is a level crossing in the excited state spectrum of the transfer
matrix (the kink in Fig.~\ref{fig:corr}) {\it en route} from the square
lattice to the isotropic triangular lattice so there are details of
the latter that the continuum theory will not reproduce very well.

Returning to the fermion action and setting $t=0$, it turns out that
in the absence of diagonal bonds we no longer need to double the unit
cell as was necessary for in Eq.~\ref{eq:pmatrix}; the fermion
operators $\psi_i$\ used in this section thus refer to the sites 
$\bf x$\ of the square lattice.

In momentum space, the resulting action is given by
\begin{equation}
S_0 = \frac12\sum_k(2i\sin k_x-2\sin k_y)
\wtpsi_{\mathbf{k}}\wtpsi_{\mathbf{-k}}
\end{equation}
where we have now reverted to the standard Euclidean field theory
convention of including an explicit minus sign with the action
in writing down the Grassmann integral. We see that the 
fermion dispersion has nodes at 
${\bf k} = (0,0)$, $(0, \pi)$, $(\pi, 0)$ and $(\pi, \pi)$. Linearizing
near these points and defining
\begin{eqnarray}
\wtchi^{1}_{\mathbf{k}} &=& \wtpsi_{\mathbf{k}} \nonumber \\
\widetilde{\bar \chi}^{1}_{\mathbf{k}} &=& 
       \wtpsi_{\pi\mathbf{\hat{y}}+\mathbf{k}} \nonumber \\
\wtchi^{2}_{\mathbf{k}} &=& 
   i \wtpsi_{\pi(\mathbf{\hat{x}} +\mathbf{\hat{y}})
                   +\mathbf{k}} \nonumber \\
\widetilde{\bar \chi}_{\mathbf{k}}^{2}
  &=& i \wtpsi_{\pi\mathbf{\hat{x}}+\mathbf{k}}
\label{eq:contfields}
\end{eqnarray}
we obtain the long wavelength form ($\partial \equiv \frac{1}{2}
\partial_x - i \partial_y$, $\bar\partial \equiv \frac{1}{2}
\partial_x + i \partial_y$)
\begin{equation}
S_0 = 2 \int d^2\mathbf{x}
\left(
\chi^{1}\bar\partial\chi^{1}+\bar\chi^{1}\partial\bar\chi^{1}
+\chi^{2}\bar\partial\chi^{2}+\bar\chi^{2}\partial\chi^{2}\right)
\end{equation}
which is a theory of two Majorana fermions, both holomorphic
($\chi^{1}$,$\chi^{2}$) and anti-holomorphic
($\bar\chi^{1}$,$\bar\chi^{2}$) fields.  The labels $1$ and $2$ here
are not the same as those used to label the two fermions in the unit
cell; now the variables $x$ and $y$ run over all sites.

The leading asymptotic behavior of
the lattice Green function  can be
recovered by inverting Eq.~\ref{eq:contfields}:
\begin{eqnarray*}
\psi({\bf x}) &=& 
\chi^{1} ({\bf x}) +(-1)^y \bar\chi^{1} ({\bf x})
\\ &&\qquad
+ i (-1)^{x+y+1}\chi^{2} ({\bf x}) + i (-1)^{x+1} \bar \chi^{2} ({\bf x}) 
\end{eqnarray*}
and using the continuum correlators 
\begin{eqnarray*}
\langle \chi^a(\bfx)\chi^a(0) \rangle&=&
\frac{1}{4 \pi} \frac{1}{ x + i y} \\
\langle \chi^a(\bfx)\chi^a(0) \rangle&=&
\frac{1}{4 \pi} \frac{1}{ x - i y}
\end{eqnarray*}
for $a=1,2$.
This procedure
yields
\begin{equation}
G({\bf x}) = \frac{1}{4 \pi} \left\{ \frac{[1 - (-1)^{x+y}]}{x + i y}
+ \frac{[ (-1)^y - (-1)^x]}{x - iy} \right\},
\end{equation}
which exhibits the two sublattice structure that characterizes the
square lattice problem. {}From this form we can recover the
asymptotics of the dimer correlations established in
Ref.~\onlinecite{Fisher63}. For example, the connected correlation
function for two horizontal dimers displaced by ${\bf x}$ is given by
the Grassmann expression,
\begin{eqnarray*}
c_{xx} ({\bf x}) &=& 
\langle \psi({\bf x}) \psi({\bf x + \hat x}) \psi({\bf 0}) 
\psi({\bf \hat x})\rangle -\\  
&&\qquad\qquad\langle \psi({\bf x}) \psi({\bf x + \hat x})\rangle
\langle \psi({\bf 0}) \psi({\bf \hat x})  \rangle \ .
\end{eqnarray*}
By Wick's theorem, $c_{xx} = - G^2({\bf x}) + G({\bf x + \hat x})
G({\bf x - \hat x})$ and using our asymptotic form for $G({\bf x})$
we find 
\begin{equation}
c_{xx} ({\bf x}) \sim \frac{(-1)^{x+y}}{4 \pi^2} \left( 
\frac{1}{x + i y} + (-1)^y \frac{1}{x - i y} \right)^2 \ ,
\end{equation}
where we have ignored constants in the denominators that generate subleading
corrections in the generic case. 
For the special case of two dimers in the same column, separated by an
even distance, this expression vanishes. However, even in that case one can 
recover the correct subleading form by reinstating the constants, as 
can be readily verified.
 
Next, we consider adding
in the diagonal bonds with fugacity $t$ and find that near the nodes
of $S_0$ it adds 
\begin{eqnarray*}
S_p &=& i t\sum_\mathbf{x}
(-1)^y\psi(\mathbf{x})\psi(\mathbf{x+\hat{x}+\hat{y}})\nonumber\\
&=&it\sum_\mathbf{k}\exp(- i k_x)\exp(-i(k_y-\pi))
\wtpsi_\mathbf{k} \wtpsi_{-\mathbf{k}+\pi\mathbf{\hat{y}}}   \ .
\end{eqnarray*}
Expanding this near the nodes of the dispersion yields
\begin{equation}
S_p =  2 i t \int d^2\mathbf{x} 
( - \bar\chi^1 \chi^1  +  \bar\chi^2  \chi^2) \ ,
\label{massterm}
\end{equation}
i.e.\ two mass terms of opposite signs. This means that the model is
invariant under the standard Kramers-Wannier duality of the Ising
model, which sends $t\to -t$. Here this just interchanges the two
fermions.  It is easy to check that our Pfaffian analysis gives
results independent of the sign of $t$. Changing $t\to -t$ leaves
invariant the Green functions $Q_s$, and $R_s$ for even $s$, while it
flips the sign of $R_s$ for odd $s$. This leaves all monomer and dimer
correlators invariant.

It follows that perturbing away from the square lattice results in a
non-analyticity in the thermodynamics, with behavior in the $d=2$
Ising universality class (with a doubling of the degrees of freedom).
Specifically, Onsager's results for the Ising model give the dimer
entropy density to be ${\cal S}(t)- {\cal S}(0) \propto - t^2 \log
t^2,$ so that the model is indeed non-analytic at $t=0$.  For $t$ near
zero, the correlation functions are now characterized by decay on the
scale of $\xi \sim 1/t$, in agreement with our earlier, exact lattice
computation (\ref{corr}).

It is interesting to ask where the order/disorder fields
($\sigma$/$\mu$) of the Ising operator algebra appear in the lattice
problem. The two known candidates here are the monomer and vison
fields, both of which exhibit an inverse square root decay on the
square lattice \cite{Fisher63,FFIM}.  In the Ising model at the
critical point, order and disorder two-point functions decay as
$|\bfx|^{-1/4}$, so the monomer and vison fields must be bilinear in
the order and disorder fields, as proven in Ref.\ \onlinecite{AYP}.  To
narrow down the correspondence, we invoke four further constraints:
(a) at $t=0$, the monomer-monomer correlator is non-zero only when the
monomers are on different sublattices\quad (b) when $t \ne 0$, monomer
correlators decay to a constant \quad (c) the vison-vison correlation
function does not exhibit any sublattice structure and decays
exponentially to zero when $t \ne 0$ \quad (d) under duality $ t
\rightarrow -t$, $\sigma$ and $\mu$ change places in each of Majorana
sectors.  Because of the opposite signs in (\ref{massterm}), one of
the sectors is in the high-temperature phase, while the other is in a
low-temperature phase. In an Ising low-temperature phase $\langle
\sigma(\bfx)\sigma(0)\rangle$ goes to a constant at large $|\bfx|$,
while in a high-temperature phase $\langle \mu(\bfx)\mu(0)\rangle$
goes to a constant. Thus for $t\ne 0$ only one of the four bilinears
decays to a constant, either $\sigma^1\mu^2$ or $\sigma^2\mu^1$, where
the superscripts refer to the two Majorana sectors.  With these we can
identify the monomer operators on the two sublattices with
$\sigma^1\mu^2 \pm i \sigma^2 \mu^1$; the relative $\pm i$ ensures
that the correlator vanishes at $t=0$ when the monomers are on the
same sublattice.  In the absence of further constraints, the vison
operator can be identified with either of the combinations $\sigma^1
\sigma^2 \pm \mu^1 \mu^2$. A fourth dimension $1/4$ operator still
needs to be identified to complete this set of identifications.

Finally, we should note that in the standard Ising model on the square
lattice, equivalent to a dimer model on Fisher's lattice, things work
somewhat differently, as we shall discuss elsewhere. \cite{fms2}  In
that problem the two Ising phases correspond to confined/deconfined
phases of the dimer problem.

\section{Concluding remarks}

We have shown that the classical model on a triangular lattice is
generically in a liquid phase with deconfined monomers. This result
contrasts with the critical correlations found on the bipartite square
and honeycomb lattices. The more complicated, non-bipartite, Fisher
lattice also exhibits a deconfined phase along with a
confinement-deconfinement transition. It would clearly be interesting
to see if other, as yet unstudied, examples follow this
classification. In this context it is worth noting that the two
critical dimer models admit height representations on account of their
bipartite character---a feature often associated with criticality in
two dimensions. It is also worth remarking that this correlation
between criticality and bipartedness connects, via the quantum dimer
model, to the Fradkin-Shenker theorem in lattice gauge
theory.\cite{fradkin79} Finally, for reasons noted in the
introduction, any further work along these lines would also be
immediately useful in advancing our understanding of quantum
magnetism.

\bigskip\bigskip
\centerline{\bf Note added}
After we submitted this paper, a preprint by A.Ioselevich, D.A.Ivanov,
M.V.Feigelman appeared (cond-mat/0206451), who studied the triangular
dimer correlations in the absence of monomers. Beyond the results reported
here, they point out that the dimer correlation length depends on
direction.

\section*{Acknowledgments}
We are grateful to Somendra Bhattacharjee and Bernard Nienhuis for
their help with the literature on dimer models, and to Chris Henley
for comments on the manuscript. RM would like to thank W. Krauth for
useful discussions. This work was supported in part by NSF
grant DMR-0104799, a DOE OJI award, and a Sloan Foundation Fellowship
(P.F.)  and by NSF grant No. DMR-9978074 and the David and Lucille
Packard Foundation (S.L.S).

\appendix
\section{Recursion relations}
\label{app:rec}
To obtain the recursion relations for the Green functions efficiently,
it is helpful to do changes several of variables.
First of all, we define $z\equiv \cos(2a)$. Since $s$ is always
odd, one can rewrite $2\cos(sa)\cos(a)=\cos((s+1)a)+ \cos((s-1)a))$
in terms of powers of $\cos(2a)$. Namely, we use
Chebyshev polynomials $T_j(y)$, which are
defined by the relation $T_j(\cos(a))=\cos(ja)$. For example,
$ T_1(y)=y$, $T_2(y)=2y^2-1$,
$T_3(y)=4y^3-3y$, and so on; a closed-form expression can be found in
\cite{Bateman}. Using this gives
\begin{equation}
Q_s = \frac{1}{2\pi}\int_{-1}^{1}  dz \frac{T_{(s+1)/2}(z)+T_{(s-1)/2}(z)}
{\sqrt{(1-z^2)(7+4z+z^2)}}
\label{Gz}
\end{equation}
This means that any $Q_s$ can be expressed as a sum of the functions
$$U_k \equiv \frac{1}{2\pi}\int_{-1}^{1}  dt\ \frac{z^k}
{\sqrt{(1-z^2)(7+4z+z^2)}}$$
for $k=1\dots(s+1)/2$.
For example, 
\begin{eqnarray*}
Q_1&=& U_1+U_0\\
Q_3&=& 2U_2 + U_1-U_0\\
Q_3&=& 4U_3-3U_2 + 2U_1-U_0\\
Q_4&=& 8U_4+4U_3-8U_2 -3U_1+U_0
\end{eqnarray*}
One also has
$$R_2=2U_1.$$

The reason we write things in terms of the $U_i$ is that it
is easy to derive recursion relations for them.
This is useful because there are
simple explicit expressions for the first few $U_k$.
Denoting the denominator of (\ref{Gz}) as $Y$, the recursion relations
follow from the fact that \cite{Bateman}
$$\frac{1}{Y}\left(mz^{m-1}Y^2+\frac{1}{2}z^m\frac{d(Y^2)}{dz}\right)
= \frac{d}{dz}\left(z^m Y\right)$$ is a total derivative. For our
case, this yields the identity
\begin{eqnarray*}
&&(k-1)U_k +2(2k-3)U_{k-1}+6(k-2)U_{k-2}\\
&&\qquad\qquad -2(2k-5)U_{k-3}-7(k-3)U_{k-4}=0.
\end{eqnarray*}

Even though the expressions for $X_s$ and $U_k$ look unwieldy,
they can be made much nicer by a simple change of variable.
Namely, defining $z=(\xi-4)/(\xi+2)$ yields
\begin{equation}
U_k = \frac{1}{4\pi}\int_1^\infty d\xi
\left(\frac{\xi-4}{\xi+2}\right)^k
\frac{1}{\sqrt{\xi^3-1}}
\end{equation}
These turn out to be related to the 
complete elliptic integral $K(k)$ with
$k=\sin(\pi/12)$, an integral whose properties were originally
investigated by Legendre. \cite{WW}
We can evaluate $U_0$ in terms of the $\beta$ function as 
\begin{eqnarray}
U_0 &=& \frac{\Gamma(7/6)}{2\Gamma(2/3)\sqrt{\pi}}\\
&=& .19326587782732139\dots
\nonumber
\end{eqnarray}
Since $Q_1=1/6$, this yields immediately that
\begin{equation}
U_1=1/6 - U_0.
\end{equation}
This then gives us our first non-trivial Green function explicitly,
because $R_2=2U_1=-.05319842232\dots$.
To utilize the recursion relations given above, we also need
$U_2$. Using various properties of the complete elliptic integrals,\cite{WW} 
we have shown that remarkably enough,
$U_2$ is simply related to the reciprocal of $U_0$, namely
\begin{equation}
U_2 = U_0 + \frac{1}{4\pi\sqrt{3} U_0} -\frac{1}{3}.
\end{equation}

\section{Derivation of the matrix for the monomer correlator}

The monomer-monomer correlator (\ref{mmdef}) is written in terms
of matrices as
$$P^{(mm)}(p)= \hbox{Pf }(M^{-1}M^{(mm)}).$$
We define
$$E \equiv M^{(mm)} - M$$ 
Since by definition the Green's function 
$G_{jk}=(M^{-1})_{jk}=-\langle \psi_j \psi_k\rangle$, we have
$$P^{(mm)}=(\det(I + GE))^{1/2}$$
where $I$ is the identity.
As seen from (\ref{mmdef}), there are three types of entries in $E$:
\begin{enumerate}
\item We need to remove the links connected to the monomer sites $1$ or $2p$.
Thus $E_{jk}=-M_{jk}$ if either $j$ or $k$ is either $1$ or $2p$.
\item We need to change the sign on the links on the string. Thus 
$E_{jk}=-2M_{jk}$ if the link $(jk)$ is part of the string.
\item We need to account for the $\psi_1\psi_{2p}$ in (\ref{mmdef}).
This is done
by $E_{2p,1}=-E_{1,2p}=i$.
\end{enumerate}

We now specialize to the case of monomers in adjacent rows a distance
$p$ apart.  Here the matrix $E$ has non-zero entries in $2(p+4)$
columns and rows.  In $E_{ab}$, the indices $r$ and $s$ run over
$1\dots p$ for the sites in the lower row, and $p+1\dots 2p$ for the
upper row. The remaining $8$ columns and rows correspond to the sites
surrounding the two monomers which are not part of the
string; we denote those adjacent to the left monomer as
$1',2',3',4'$, and those adjacent to the right monomer as $5',6',7',8'$.

Since we evaluated all the Green functions in section III for
$t=1$, at this point we could just plug in all the numbers, multiply
the matrices, evaluate the determinant, and take the square root to
get the monomer correlators.  However, the determinant can be
simplified a great deal, and is quite elegant. In fact, we can rewrite
the correlator as a determinant of a matrix half the size.

Let us first consider the first column of $I+GE$. The non-zero entries
of $E_{b1}$ are when $b=2p$ or $b=2,p+1,1',2',3',4'$ (the latter sites
surrounding the site $1$). Note
that $\sum_c G_{ac}M_{c1}=\delta_{a1}$ when $c$ is summed over the sites
around $1$. Then
$$G_{ab}E_{b1}=iG_{a,2p}-\sum_c G_{ac}M_{c1}= G_{a,2p}-\delta_{a1}$$
Thus 
$$(I+GE)_{a1}=iG_{a,2p}.$$ 
Similarly, $(I+GE)_{a,2p}=-iG_{a1}$.
We now consider the columns coming from the sites adjacent to site $1$
but not part of the string. For these sites $c'=1',2',3',4'$, we have
$E_{ab}\propto\delta_{a1}$. Thus in the $k$th column of $I+GE$ we have
$G_{ab}E_{bc'}=G_{a1}E_{1c'}$. However, we can essentially remove
this column without affecting the determinant. If we add the $2p$th
column times $iE_{1c'}$ to the $c'$th column, the $c'$th column of
$(I+GE)$ is just $\delta_{ac'}$. Thus the only non-zero entry in the
$c'$th column is on the diagonal, and is $1$. The determinant with
these rows and columns removed is thus identical to those with these columns
present. The rows and columns $5',6',7',8'$ are removed in a similar manner.

We denote the modified $2p\times 2p$ matrix 
(with the same Pfaffian) as ${\cal M}$.
We can continue with such column manipulations to simplify the matrix
further. Consider the $p$th column. We have
$$(I+GE)_{ap}=\delta_{ap} + 2iG_{a,2p-1} - it G_{a,2p}.$$
The last piece can be removed without changing the determinant by 
adding $t$ times the first column of ${\cal M}$ 
to this column. 
This gives 
$${\cal M}_{ap}=\delta_{ap} + 2iG_{a,2p-1}.$$
Now consider the
$p-1$th column:
$$(I+GE)_{a,p-1}=\delta_{a,p-1} + 2iG_{a,2p-2} - 2it G_{a,2p-1}.$$
We can simplify this by adding the $t$ times the
$p$th column to this, yielding
for the modified matrix
$${\cal M}_{a,p-1}=\delta_{a,p-1} +t \delta_{ap}+2iG_{a,2p-2}.$$
Continuing in this fashion gives
$${\cal M}_{ab}=\sum_{j=b}^p \delta_{aj}t^{a-b} +2iG_{a,b+p-1}.$$
for $b=2\dots p$ and $a=1\dots 2p$.
We already have
$${\cal M}_{a1}=iG_{a,2p}.$$
We can do the analogous manipulations for the columns $c=p+1\dots 2p-1$.
This yields 
$${\cal M}_{ac}=\sum_{j=p+1}^c \delta_{aj} t^{c-a}-2iG_{a,c-p+1}.$$
and
$${\cal M}_{a,2p}=-iG_{a,1}.$$

We can now write ${\cal M}$ in terms of the $Q_s$ and $R_s$ defined
in the last section. For example, 
$$
{\cal M}_{a,2p}=
\begin{cases}
i(-1)^{[a/2]} Q_{1-a} & 1\le a\le p\\
(-1)^{[(a-p-1)/2]} R_{a-p}&p+1\le a \le 2p
\end{cases}
$$
where $[x]$ is the greatest integer less than $x$.
Putting this all together yields, for example, for $p=4$
\begin{widetext}
$$
{\cal M}=
\left [\begin {array}{cccccccc} -R_{{4}}&2\,R_{{1}}&2\,R_{{2}}&-2\,R_{
{3}}&2\,iQ_{{1}}&0&-2\,iQ_{{3}}&0\\\noalign{\medskip}-R_{{3}}&1-2\,R_{
{0}}&2\,R_{{1}}&2\,R_{{2}}&0&2\,iQ_{{1}}&0&-iQ_{{1}}
\\\noalign{\medskip}R_{{2}}&t-2\,R_{{-1}}&1-2\,R_{{0}}&2\,R_{{1}}&-2\,
iQ_{{1}}&0&2\,iQ_{{1}}&0\\\noalign{\medskip}R_{{1}}&t^2+2\,R_{{-2}}&t-2
\,R_{{-1}}&1-2\,R_{{0}}&0&-2\,iQ_{{1}}&0&iQ_{{3}}\\\noalign{\medskip}i
Q_{{3}}&0&-2\,iQ_{{1}}&0&1-2\,R_{{0}}&t-2\,R_{{-1}}&t^2+2\,R_{{-2}}&R_{{
1}}\\\noalign{\medskip}0&2\,iQ_{{1}}&0&-2\,iQ_{{1}}&2\,R_{{1}}&1-2\,R_
{{0}}&t-2\,R_{{-1}}&R_{{2}}\\\noalign{\medskip}-iQ_{{1}}&0&2\,iQ_{{1}}
&0&2\,R_{{2}}&2\,R_{{1}}&1-2\,R_{{0}}&-R_{{3}}\\\noalign{\medskip}0&-2
\,iQ_{{3}}&0&2\,iQ_{{1}}&-2\,R_{{3}}&2\,R_{{2}}&2\,R_{{1}}&-R_{{4}}
\end {array}\right ]
$$
where we have used the facts that $Q_{-s}=Q_{s}$ and $Q_{s}=0$ for $s$ even.
This matrix looks much nicer if we reshuffle rows and columns. We
permute the first column through the others so that it becomes the
$p$th column, and permute the $2p$th column through so that it becomes
the $p+1$th column. This does not change the determinant. 
We then relabel the indices $p+1\dots 2p$ in reverse order
(i.e. interchange $2p\leftrightarrow p+1$, $2p-1\leftrightarrow p+2$, etc.)
We also multiply the $p$th and the $p+1$th columns by 2,
so that we need to divide the resulting determinant by 4.
These manipulations put the matrix into the form
\begin{equation}
\begin{pmatrix}
{\cal R}&{\cal Q}\\{\cal Q}&{\cal R}
\end{pmatrix}
\label{block}
\end{equation}
where ${\cal R}$ and ${\cal Q}$ are the $p\times p$ matrices
defined by 
\begin{eqnarray}
{\cal R}_{ij}&=&(-1)^{[(j-i)/2]}2R_{j-i+1} + \theta(i-j)t^{i-j-1}\\
{\cal Q}_{ij}&=&i(-1)^{[(j+i)/2]}2Q_{p+1-i-j} 
\end{eqnarray}
where $\theta(x)=1$ for $x>0$ and
$0$ for $x\le 0$. For example, for $p=6$ we have
$$
{\cal R}=
\begin{pmatrix}
2R_1&2R_2&-2R_3&-2R_4&2R_5&2R_6\\
1-2R_0&2R_1&2R_2&-2R_3&-2R_4&2R_5\\
t-2R_{-1}&1-2R_0&2R_1&2R_2&-2R_3&-2R_4\\
t^2+2R_{-2}&t-2R_{-1}&1-2R_0&2R_1&2R_2&-2R_3\\
t^3+2R_{-3}&t^2+2R_{-2}&t-2R_{-1}&1-2R_0&2R_1&2R_2\\
t^4-2R_{-4}&t^3-2R_{-3}&t^2+2R_{-2}&t-2R_{-1}&1-2R_0&2R_1\\
\end{pmatrix}
$$
$$
{\cal Q}=
\begin{pmatrix}
2iQ_5&0&-2iQ_3&0&2iQ_1&0\\
0&-2iQ_3&0&2iQ_1&0&-2iQ_1\\
-2iQ_3&0&2iQ_1&0&-2iQ_1&0\\
0&2iQ_1&0&-2iQ_1&0&2iQ_3\\
2iQ_1&0&-2iQ_1&0&2iQ_3&0\\
0&-2iQ_1&0&2iQ_3&0&-2iQ_5\\
\end{pmatrix}
$$
The determinant of matrices of the form (\ref{block}) can be 
simplified:\cite{Fowler}
$$
\det \begin{pmatrix}
{\cal R}&{\cal Q}\\{\cal Q}&{\cal R}
\end{pmatrix}
=\det
\begin{pmatrix}
{\cal R}+{\cal Q}&{\cal Q}\\{\cal Q}+{\cal R}&{\cal R}
\end{pmatrix}
=\det
\begin{pmatrix}
{\cal R}+{\cal Q}&{\cal Q}\\0&{\cal R}-{\cal Q}
\end{pmatrix}
=\det({\cal R}+{\cal Q}) \det({\cal R}-{\cal Q}) .
$$
\end{widetext}
In our case, the latter two determinants are the same, because ${\cal
R}$ is symmetric around the off-diagonal (the diagonal from upper
right to lower left), while ${\cal Q}$ is antisymmetric around the
off-diagonal. 
Thus
$$\det{{\cal M}} = \frac{1}{4}\left( \det({\cal R}+{\cal Q}) \right)^2.$$
This yields the result (\ref{eq:monodet})
for the monomer-monomer correlator above.


\begin{thebibliography}{99}


\bibitem{kasteleyn}
P.~W. Kasteleyn, \tit{Physica}{27}{1209}{1961}{The statistics of dimers
on a lattice}; \tit{\jmp}{4}{287}{1963}{Dimer statistics and 
phase transitions}


\bibitem{fishdimis} 
M.~E. Fisher, \tit{\jmp}{7}{1776}{1966}{On the
Dimer Solution of Planar Ising Models}


\bibitem{NYB} For a review, see J.~F. Nagle, C.~S.~O. Yokio and
S.~M. Bhattacharjee, ``Dimer models on anisotropic lattices'', in {\it
Phase Transitions and Critical Phenomena}, ed.~by C.~Domb and
J.~Lebowitz, v.~13 (Academic Press, 1980).


\bibitem{Fisher63}
M.~E. Fisher and J. Stephenson, \tit{\pr}{132}{1411}{1963}{Statistical 
Mechanics of Dimers on a Plane Lattice. II. Dimer 
Correlations and Monomers}


\bibitem{MW} B.M. McCoy and T.T. Wu, {\it The Two-Dimensional Ising Model}
(Harvard, 1973)


\bibitem{Anderson87}
P.~W. Anderson, 
\tit{Science}{235}{1196}{1987}{THE RESONATING VALENCE
BOND STATE IN LA2CUO4 AND SUPERCONDUCTIVITY}.

\bibitem{Rokhsar88}
D.~S. Rokhsar and S.~A. Kivelson,
\tit{\prl}{61}{2376}{1988}{SUPERCONDUCTIVITY AND THE QUANTUM HARD-CORE
DIMER GAS}.
 
\bibitem{kenyon} The precise
connection between classical and quantum correlations requires additional
assumptions about the ergodicity of the quantum dimer Hamiltonian in the
space of dimer configurations \cite{Rokhsar88,MStrirvb}. This problem has been
addressed for a finite triangular lattice with open boundary conditions in
C. Kenyon and E. Remila, \tit{Discrete Math.}{152}{191}{1996}{Perfect
matchings in the triangular lattice}.


\bibitem{MStrirvb}
R. Moessner and S.~L. Sondhi, 
\tit{\prl}{86}{1881}{2001}{Resonating Valence 
Bond Phase in the Triangular Lattice Quantum Dimer Model}.

\bibitem{Wu}
F.~Y. Wu, \tit{Phys.~Rev.}{168}{539}{1967}{Remarks on the Modified
Potassium Dihydrogen Phospate Model of a Ferroelectric}

\bibitem{wen}  X. G. Wen,
\tit{\prb}{40}{7387}{1989}; \tit{\ijmpb}{4}{239}{1990};
X. G. Wen and Q. Niu,
\tit{\prb}{41}{9377}{1990}{GROUND-STATE DEGENERACY
OF THE FRACTIONAL QUANTUM HALL STATES IN THE
PRESENCE OF A RANDOM POTENTIAL AND ON HIGH-GENUS RIEMANN SURFACES}.

\bibitem{duality}
R. Moessner, S.~L. Sondhi and E. Fradkin,
\tit{\prb}{65}{024504}{2002}{Short-ranged RVB physics, 
quantum dimer models and Ising gauge theories}

\bibitem{nagle}
J.~F. Nagle, \tit{\pr}{152}{190}{1966}{New Series-Expansion Method
for the Dimer Problem}.

\bibitem{gaunt}
D.~S. Gaunt, \tit{\pr}{179}{174}{1969}{Exact Series-Expansion
Study of the Monomer-Dimer Problem}.

\bibitem{samuel}
S. Samuel, \tit{\jmp}{21}{2806}{1980}{The use of 
anticommuting variable integrals in statistical mechanics. I. 
The computation of partition functions.}

\bibitem{mila}
F. Mila, \tit{\prl}{81}{2356}{1998}{Low-Energy Sector of the 
S  = 1/2 Kagome Antiferromagnet}

\bibitem{wolffzitt}
W.~F. Wolff and J. Zittartz, \jour{\zpb}{49}{139}{1982}.

\bibitem{fn-polyaloop} An actual magnetic model will exhibit a
finite density of spinons as a finite temperature. Our diagnostic
is the equivalent of the ``Polyakov loop'' for gauge theories
{\it without} dynamical matter, at finite temperatures. The latter
measures the free energy of two static quarks at varying separations.

\bibitem{fn-werner} This problem was concurrently studied numerically
(by Monte Carlo simulations and by exact enumeration) 
by W. Krauth and R. Moessner, cond-mat/0206177.

\bibitem{fisherdimer}
M.~E. Fisher, \tit{\pr}{124}{1664}{1961}{Statistical Mechanics 
of Dimers on a Plane Lattice}.

\bibitem{temperley}
M.~E. Fisher and H.~N.~V. Temperley, \jour{Phil.\ Mag.}{6}{1061}{1961}

\bibitem{fn-sign} 
The corresponding expression for $g$\ in
Ref.~\onlinecite{MStrirvb} contains an error in the sign of the argument of 
the first
exponential, which is corrected here.

\bibitem{Hartwig} R.~E. Hartwig, \tit{\jmp}{7}{286}{1966}{Monomer Pair
Correlations}

\bibitem{slowhole}
R. Moessner and S.~L. Sondhi,
\tit{\prb}{62}{14122}{2000}{Slow holes in the triangular Ising 
antiferromagnet}

\bibitem{AYP} H. Au-Yang and J.~H.~H. Perk, 
\tit{Phys.~Lett.~A}{104}{131}{1984}{Ising Correlations at the 
Critical Temperature}

\bibitem{isc} We discovered this using the Inverse Symbolic Calculator
at http://www.cecm.sfu.ca/projects/ISC/ . We thank Michael Stone for
telling us about this resource.

\bibitem{ReadCha} N. Read and B. Chakraborty,
\tit{\prb}{40}{7133}{1989}{STATISTICS OF THE EXCITATIONS
OF THE RESONATING-VALENCE-BOND STATE}.

\bibitem{Kivelson}
S. Kivelson, \tit{\prb}{39}{259}{1989}{Statistics of holons
in the quantum hard-core dimer gas}.

\bibitem{bfg} L. Balents, M.P.A. Fisher and S.M. Girvin,
Phys.~Rev.~B {\bf 65}, 224412 (2002) [cond-mat/0110005].


\bibitem{shtengel} K. Shtengel, private communication

\bibitem{FFIM} G. Forgacs, \tit{\prb}{22}{4473}{1980}{Ground-state 
correlations and universality in two-dimensional fully frustrated 
systems}.

\bibitem{fms2} R.~Moessner and S.~L. Sondhi, cond-mat/0212363.

\bibitem{fradkin79}
E.~H.~Fradkin and S.~H.~Shenker,
Phys.\ Rev.\ D {\bf 19}, 3682 (1979).

\bibitem{Bateman} A.~Erdelyi, {\it Bateman Manuscript Project: 
Higher Transcendental Functions, vol.~2} (McGraw-Hill, 1953)

\bibitem{WW} E.~T. Whitaker and G.~N. Watson, {\it A Course in Modern
Analysis} (Cambridge, 1927)

\bibitem{Fowler} We thank Michael Fowler for providing this proof.

\end{thebibliography}
\end{document}